%% file: shearPeaks.tex
\documentclass[iop]{emulateapj}

\usepackage{amsmath}

\newcommand{\BE}{\begin{equation}}
\newcommand{\EE}{\end{equation}}
\newcommand{\BA}{\begin{eqnarray}}
\newcommand{\EA}{\end{eqnarray}}
\def\kipac{KIPAC, Stanford University, 452 Lomita Mall, 
Stanford, CA 94309, USA}

\def\ssl{Space Sciences Laboratory, University of California, 
 Berkeley, CA 94720, USA}
\def\purdue{Department of Physics, Purdue University, 
 West Lafayette, IN 47907, USA}
\def\uw{Department of Astronomy, University of Washington, 
Seattle, WA 98195}

\def\miami{Department of Physics, University of Miami, Coral Gables, FL, 33124, USA}
\def\brookhaven{Physics Department, Brookhaven National Laboratory, Upton, NY 11973, USA}
\def\columbia{Department of Astronomy and Astrophysics, Columbia University, New York, NY 10027, USA}

\shorttitle{Shear Peak Statistics}
\shortauthors{Bard et al.}

\begin{document}

\title{Effect of Measurement Errors on Predicted Cosmological Constraints from Shear Peak Statistics with LSST. }

\author{D. Bard\altaffilmark{1}, J. M. Kratochvil\altaffilmark{2}, C. Chang\altaffilmark{1}, M. May\altaffilmark{3}, S. M. Kahn\altaffilmark{1}, \\ 
Y. AlSayyad\altaffilmark{4}, Z. Ahmad\altaffilmark{5}, J. Bankert\altaffilmark{5}, A. Connolly\altaffilmark{4}, R. R. Gibson\altaffilmark{4}, K. Gilmore\altaffilmark{1}, E. Grace\altaffilmark{5}, \\
Z. Haiman\altaffilmark{6}, M. Hannel\altaffilmark{5}, K. M. Huffenberger\altaffilmark{2}, J. G. Jernigan\altaffilmark{6}, L. Jones\altaffilmark{4}, S. Krughoff\altaffilmark{4}, S. Lorenz\altaffilmark{5}, \\
 S. Marshall\altaffilmark{1}, A. Meert\altaffilmark{5}, S. Nagarajan\altaffilmark{5}, E. Peng\altaffilmark{5}, J. Peterson\altaffilmark{5}, A. P. Rasmussen\altaffilmark{1}, M. Shmakova\altaffilmark{1}, \\
N. Sylvestre\altaffilmark{5}, N. Todd\altaffilmark{5}, M. Young\altaffilmark{5} }
\email{djbard@slac.stanford.edu}
\altaffiltext{1}{\kipac}
\altaffiltext{2}{\miami}
\altaffiltext{3}{\brookhaven}
\altaffiltext{4}{\uw}
\altaffiltext{5}{\purdue}
\altaffiltext{6}{\columbia}
\altaffiltext{7}{\ssl}

\begin{abstract}

The statistics of peak counts in reconstructed shear maps contain information beyond the power spectrum, and can improve cosmological constraints from measurements of the power spectrum alone if systematic errors can be controlled. 
We study the effect of galaxy shape measurement errors on predicted cosmological constraints from the statistics of shear peak counts with the Large Synoptic Survey Telescope (LSST). 
We use the LSST image simulator in combination with cosmological N-body simulations to model realistic shear maps for different cosmological models. 
We include both galaxy shape noise and, for the first time, measurement errors on galaxy shapes. 
We find that the measurement errors considered have relatively little impact on the constraining power of shear peak counts for LSST.

\end{abstract}


\input{introduction}

\input{formalism}

\input{simulations}

\input{galaxyModel}

\input{peakcounts}

\input{analysis}

\input{cosmology}

\input{summary}

\acknowledgments

We would like to thank Brandon Calabro for useful discussions on shear maps and aperture mass statistics. J.M.K. and K.M.H. receive support from NASA's Jet Propulsion Laboratory subcontract 1363745. This research utilized resources at the New York Center for Computational Sciences, a cooperative effort between Brookhaven National Laboratory and Stony Brook University, supported in part by the State of New York. This work is supported in part by the U.S. Department of Energy under Contract No. DE-AC02-98CH10886.

\end{document}

%% file: introduction.tex
\section{Introduction}
\label{sec:intro}

Weak gravitational lensing (WL) by large-scale cosmic structures has emerged as one of the most promising methods to constrain the
parameters of both dark energy and dark matter (e.g. \citet{DETF}; see also \citet{HJ08,Munshi+08} for reviews).  
The COSMOS survey has provided independent evidence of the accelerated expansion of the Universe from cosmological WL measurements \citep{COSMOS, Semboloni+11b}. 
Over the next decade, the Large Synoptic Survey Telescope (LSST) and other large surveys covering several thousand square degrees will produce galaxy catalogues of unprecedented quality.\footnote{Forthcoming and planned surveys include those
by LSST (www.lsst.org), by Hyper Suprime-Cam
(HSC, www.naoj.org/Projects/HSC/HSCProject.html),
the Dark Energy Survey (DES,
www.darkenergysurvey.org), the Kilo-Degree Survey
(KIDS, http://kids.strw.leidenuniv.nl), Pan-STARRS
(http://pan-starrs.ifa.hawaii.edu/public), and Euclid
(http://sci.esa.int/euclid).
}
These surveys will provide WL datasets with an enormous wealth of information about structure formation, enabling not just the study of traditional two-point statistics like the power spectrum, but also the extraction of information on non-Gaussianity. 
The combination of these measurements  will help substantially tighten the constraints of cosmological parameters.

Galaxy clusters are collapsed objects that provide a complementary probe of cosmology to the power spectrum. 
Indeed, the cluster mass function has long been considered a useful probe of cosmological models. 
It can be approximated analytically, and has a strong dependence on the cosmological parameters $\Omega_m$, $\sigma_8$ and $w$, where  $\Omega_m$ is the fractional matter density of the universe,  $\sigma_8$ is the normalization of the matter power spectrum at the length scale $8h^{-1}$Mpc and  $w$ is the evolution of the equation of state of dark energy. 
Clusters can be identified and their masses measured through several different techniques, including x-ray observations, the SZ effect, galaxy counts and weak gravitational lensing ~\citep[see][for an overview]{allen11}. 
The measurement of cluster mass using WL has the advantage that it is independent of the luminous and dynamic properties of the galaxies, and is sensitive to both the baryonic and dark matter components. 
However using WL to {\it detect} clusters, by searching for peaks in reconstructed lensing maps, is more problematic. 
Shear peaks detected in WL surveys are often not due to single galaxy clusters, but to chance alignments of structure along the line-of-sight~\citep{ham04}. 
In addition, genuine clusters that are aligned with matter underdensities along the line-of-sight can be missed. 
Cluster counts using WL alone therefore tend to have low completeness and low purity~\citep{ham04}.

Peaks in WL maps are a direct observable in WL surveys and can constrain cosmology, regardless of whether they originate from a single galaxy cluster  or a random superposition of matter over-densities. 
While harder to predict theoretically than the cluster mass function, they are observationally cleaner with fewer opportunities for systematic errors to complicate the interpretation of the measurement. 
Contributions from filaments and other chance alignments encode additional information about the structure of matter beyond the cluster mass function, making peak counts a probe of cosmological parameters complementary to measurements from cluster counts.

In the past several years, there has been a significant increase in interest in lensing peaks and other closely related statistics.\footnote{To our knowledge, lensing peaks were first considered as a probe of cosmology in the early ray-tracing simulations by \cite{JV00}, who studied the $\Omega_m$--dependence of the peak counts.} 
Most work has concentrated on peak counts in maps of convergence, which are easy to simulate but observationally harder to reconstruct than maps of reduced shear (see Section~\ref{sec:formalism} for definitions of convergence and reduced shear). 
\citet{JSW00} studied the probability distribution function of the convergence and \citet{whm09} investigated its cumulative version, the fractional area of ``hot spots'' on convergence maps. 
Both statistics are similar to peak counts in the high-convergence limit and have been shown to have useful cosmology sensitivity. 
The fractional area statistic is also known as $V_0$, one of the three Minkowski functionals for two-dimensional thresholded fields. Minkowski functionals are related to peaks and had been proposed as a weak-lensing statistic by \citet{Sato:2001cb} and \citet{g02}. 
More recently, \citet{Maturi:2009as} constructed an analytical approximation to the $V_2$ Minkowski functional, which is the genus statistic and also corresponds to peak counts in the high-threshold limit.  
The full set of Minkowski functionals in the context of WL has been studied extensively both theoretically~\citep{Munshi:2011wu} and in ray-tracing simulations~\citep{Minkowski}.  
In a different approach, peak counts have also been studied in wavelet space~\citep{Pires_WPC:2009}, and found to break the degeneracy in $(\sigma_8, \Omega_m)$ cosmological models found in measurements of the power spectrum alone.

Preliminary studies \citep{Marian:2008fd, Marian:2009wi} that defined peaks as local density maxima were based on 2D projections of the 3D mass distribution in low-resolution N-body simulations.  
Weak-lensing peak counts using ray-traced simulations were subsequently studied by~\citet{Dietrich:2009jq} and \citet{Kratochvil:2009wh} and more recently in~\citet{Li2011} and \citet{Marian:2011rg}. 
Based on simulations with better mass resolution, these references revealed that low--amplitude peaks (which typically do not correspond to single collapsed dark matter halos) contain more cosmological information than high-amplitude peaks. It should be noted that the range of peak heights qualifying peaks as \lq\lq low\rq\rq\ varies greatly between these references.

Several other aspects of WL peak counts have also been explored. 
WL peaks were used by \citet{Marian:2010mh} and \citet{Maturi:2011am} to predict constraints on the primordial non-Gaussianity parameter $f_{\mathrm{NL}}$. \citet{Yang:2011zz} studied the origin of the cosmologically important low peaks, and found that they are typically caused by a combination of 4--8 low-mass halos. \citet{Minkowski} and \citet{Marian:2011rg} demonstrated that cosmological constraints from peaks can be tightened by combining several angular smoothing scales. \citet{Pires_PC:2012} compared WL peak counts directly to two other commonly used non-Gaussian statistics, skewness and kurtosis, and found the peak counts to be superior in information extraction from WL maps. Finally, \cite{vanderplas12} studied the effect of masks on shear peak counts and showed that using Karhunen-Lo\`eve analysis can mitigate biases on peak count distributions caused by masked regions, and  can also reduce the number of noise peaks. A comprehensive study of the uncertainty that the presence of masks introduces into peak counts in real observational situations has yet to be performed. 

Previous work on this subject has therefore determined the value of peak counts in constraining cosmology, both alone and in combination with other lensing measurements, where peak counts can break degeneracies in cosmological parameter estimation. 
Attention has also been paid to optimizing the extraction of cosmological information. 
Work by~\citet{Maturi:2009as} and~\citet{Marian:2010mh} has concentrated on determining the optimal filter size and shape, and how filters of different sizes can be combined to increase the information extracted from shear maps. 
\citet{Dietrich:2009jq} demonstrated the value in redshift-dependent measurement of shear peaks. 
However, to date there has been no effort to include measurement errors in the predictions made from the above simulations.

This paper introduces a framework to produce realistic galaxies that can be used to trace the shear maps produced in cosmological simulations, with sizes, magnitudes, redshifts and signal-to-noise properties matching observed distributions, and measurement errors matching expected uncertainties from a ten-year LSST survey. 
We take a forward-modeling approach where we compare a dataset produced with a particular cosmology with datasets produced for other cosmological models. 
In this way, we can compare the expected results from the different cosmologies and, eventually, determine the best-fit to the data. 

This paper is organized as follows. 
The WL formalism and aperture mass calculation are introduced in Section~\ref{sec:formalism}. 
To calculate these shear peak statistics, we start with a large suite of N-body simulations described in ~\citet{Minkowski} and \citet{Yang:2011zz}, which are ray-traced in order to obtain maps of the shear and convergence parameters for seven different cosmological models, covering the cosmological parameter space in  $\Omega_m$, $\sigma_8$ and $w$. 
These are described in Section~\ref{sec:sims}. 
We seed these shear maps with source galaxies in order to obtain a mock dataset for each cosmological model. 
Realistic galaxies are essential to predicting realistic constraints on cosmology, so we must include all effects that will impact the quality of the measurements. 
Measurement uncertainties depend largely on the signal-to-noise ratio of the flux of the galaxy, with fainter galaxies having lower signal-to-noise. 
We therefore use the LSST Image Simulator~\citep{imsim} input catalogues to identify the intrinsic properties to be used for the source galaxies, such as size, magnitude and redshift. 
Uncertainties in shape measurement are determined from a large suite of LSST simulations, used to model the expected errors due to atmospheric and instrumental effects and the residual contributions to galaxy shape distortion after the ellipticity of the point spread function (PSF) has been interpolated to the galaxy position and deconvoluted from the galaxy shape. 
This process is described in Section~\ref{sec:gal_mod}, and provides a mock catalogue of galaxy shapes representative of that which would be obtained after 10 years of LSST data, for each of the seven cosmological models. 
We then calculate the aperture mass over each of these mock catalogues, and look for peaks in the maps of signal-to-noise ratio (SNR) of the aperture mass statistic. 
The resulting peak counts are described in Section~\ref{sec:peakcounts}, which allow us to make the first realistic predictions of constraints on cosmological parameters from shear peak statistics for LSST. 
We describe the process by which we calculate the constraints on cosmology by comparing the different mock datasets in Section~\ref{sec:analysis}, and discuss our results in Section~\ref{sec:cosmo}. 
Finally, we summarize our work in Section~\ref{sec:summary}.

%% file: formalism.tex
\section{Formalism}
\label{sec:formalism}
Photons from distant galaxies are deflected by the tidal gravitational field of matter along the line of sight. 
If the lensed image of a galaxy is smaller than the characteristic scale of the lensing potential, the distortion of the galaxy shape can be described by a linearized lens mapping, given by the Jacobian
\begin{equation*}
A = (1 - \kappa) \left( \begin{array}{cc} 1-g_1 & -g_2 \\ -g_2 & 1+g_1 \end{array} \right), 
\end{equation*}
where $g$ is the reduced shear $g = \frac{\gamma}{1-\kappa}$. 
The complex shear $\gamma \equiv \gamma_1 + i\gamma_2$ describes the distortion of the galaxy shape, and the convergence, $\kappa$, describes the magnification of the galaxy image relative to its source. 
For a full derivation of these parameters, see for example~\citet{bs}. 

Of course, we cannot measure the shear parameters directly, but must estimate them from the resulting small distortions in observed galaxy shapes. 
We parametrize galaxy shapes by the complex ellipticity $\epsilon = \epsilon_1 + i \epsilon_2$, where the components of ellipticity are normalized moments of the light intensity of the object $I_{i,j}$ weighted by a Gaussian function $W(x_1, x_2)$:
\begin{eqnarray*}
\epsilon_1 = \frac{I_{11} - I_{22}}{I_{11} + I_{22}}, \hspace{10pt} \epsilon_2 = \frac{2I_{12}}{I_{11} + I_{22}}, \hspace{50pt} \\
I_{ij} = \frac{ \int \int W(x_1, x_2) f(x_1, x_2) x_i x_j dx_1 dx_2} {\int \int W(x_1, x_2) f(x_1, x_2) dx_1 dx_2}, \hspace{10pt} i, j = 1,2. 
\end{eqnarray*}

The observed galaxy ellipticity is a combination of the intrinsic galaxy ellipticity $\epsilon_{int}$ and reduced shear $g$: $\epsilon_{obs} = \epsilon_{int} + g$. 
Shape noise from the intrinsic ellipticity of galaxies $\sigma^2_{int} = \langle \epsilon^2_{int}\rangle$ is much larger than $g^2$, so to obtain $g$ we can average over large numbers of galaxies (assuming that galaxy shapes and orientations are random over a large enough area of the sky). 
In this case, the observed ellipticity $\langle \epsilon_{obs}\rangle = \langle g \rangle$. 
The uncertainty in a measurement of $g$, $\sigma_g$, is therefore a combination of the galaxy shape noise and measurement uncertainty $\sigma^2_g = \sigma^2_{int} + \sigma^2_{meas}$~\footnote{We define ``error'' as the residual between the measured and true quantity, and ``uncertainty'' as the standard deviation of the differences between the measured and true quantity.}. 
Previous work \citep[e.g.][]{Dietrich:2009jq, Maturi:2009as, Marian:2010mh} has considered the impact of the shape noise $\sigma_{int}$ but not the measurement uncertainty $\sigma_{meas}$. 

Matter over-densities along the line of sight will cause the shear field, and therefore the observed shapes of galaxies, to be tangentially aligned around the projected peak of the over-density. 
We can use this property of shear fields to reconstruct the aperture mass, which is a weighted sum over the tangential components of galaxy shapes around a point. 
We define the aperture mass at position $\theta_0$ as in~\citet{schneider06},
\begin{equation*}
M_{ap}(\theta_0) = \int d^2 \boldsymbol{\theta} Q(\theta) g_t(\theta, \theta_0),
\end{equation*}
where $g_t$ is the tangential component of reduced shear relative to $\theta_0$ defined as 
\begin{equation*}
g_{\mathrm{t}}(\theta, \theta_0)=-(g_1\cos(2\phi)+g_2\sin(2\phi)).
\end{equation*}
$\phi$  is the angle with respect to the horizontal axis between positions $\theta_0$ and $\theta$ in the map. 
Note the minus sign, and the factor of two in the angles (necessary because shear is spin-2, not a vector). 
$Q(\theta)$ is the weighting function, and determines the statistical properties of $M_{ap}$. 
In practice, the shear field is sampled by galaxies and we measure the reduced shear of these galaxies.  
We  therefore estimate the aperture mass by summing over the tangential components of galaxy shapes using 
\begin{equation}
M_{ap}(\theta_0) = \frac{1}{N_g} \sum^{N_g}_{i=1}  Q(\theta) g_{i,t}.
\label{eqn:map}
\end{equation}

If the weight function $Q(\theta)$ follows the expected shear profile of a mass peak then the aperture mass is a matched filter for detecting mass peaks. 
We use the spherically symmetric function introduced by ~\citet{schirmer07}, which follows an NFW~\citep{NFW} profile with exponential cutoffs as $x \to 0$ and $x \to \infty$:
\begin{equation*}
Q_{NFW}(x, x_c) = \frac{1}{ 1 + e^{6-160x} + e^{-47+50x}} \frac{\tanh(x/x_c)}{x/x_c}. 
\end{equation*}
Here, $x=\theta_i/\theta_{max}$, where $\theta_{max}$ gives the radius to which the filter is tuned. 
We use a value of 5.6$arcmin$. 
$x_c$ is a constant, set to 0.15, which has been empirically determined to be a good value for shear peak counting~\citep{hetterscheidt05}.\footnote{However, it is not yet clear whether this value or this filter shape in general is the best choice for the low shear peaks, which have been discovered to be cosmologically important \citep{Dietrich:2009jq, Kratochvil:2009wh} and been shown to be due to projections of multiple clusters \citep{Yang:2011zz} since the publication of \citet{hetterscheidt05}. }
The rms dispersion of $M_{ap}$ in the case of no lensing is determined from the dispersion of the intrinsic shape noise of galaxies~\citep{bs},
\begin{equation}
\sigma(M_{ap}) = \frac{\sigma_{g}}{\sqrt{2}n} \sqrt{ \sum_i Q^2(\theta_i)}.
\label{eqn:sigmap}
\end{equation}
Providing the lensing is weak within the radius of the aperture, $\sigma(M_{ap})$ will be close to the rms dispersion in the presence of lensing. 
It can therefore be used as an estimate of the uncertainty of the aperture mass. 
We can calculate the noise directly from the data, and look for peaks in the map of SNR, 
\begin{equation}
\mathrm{SNR}(\theta_0) = \frac{\sqrt{2}\sum_i Q(\theta_i)g_{i,t}} {\sqrt{ \sum_i Q^2(\theta_i)g_i^2}}.
\label{eqn:snr}
\end{equation}
We use a pixel size of 12.2$\arcsec^2$ for this map. 
We define peaks in the SNR map as all pixels in the map above a certain threshold having 8-connectivity (i.e. pixels which are connected along the sides or by the corners). While there are other possible definitions of peaks, this one---corresponding to the definition of local maxima in a pixelized map---is simple and  makes few assumptions about the underlying nature of the peaks. 

We are working with thousands of 12 deg$^2$ simulated shear and convergence maps, with each map containing $\sim$ 1.5 million galaxies. 
Calculating the aperture mass for all maps is a significant computational problem, which we address by taking advantage of the properties of graphics processing units (GPUs). 
The implementation of the aperture mass calculation on the GPU is described in detail in~\citet{gpu}. 
By using the GPU we can reduce the computation time per map from several hours to a few minutes.

Previous work~\citep{Yang:2011zz} has determined that peak counts in convergence maps contain additional information not provided by the power spectrum alone. 
In order to make a similar determination about the information in peak counts in reduced shear maps, we must also calculate the power spectrum using a simple Fourier transformation. 
We will also use this information to constrain cosmological parameters, alone and in combination with the peak counts, as described in Section~\ref{sec:cosmo}.

%% file: simulations.tex
\section{Simulations}
\label{sec:sims}
In order to predict peak counts from different cosmological models we must use a large suite of N-body simulations representing these models, ray-traced to produce shear maps. The large-scale structure simulations and shear maps we use in this analysis were created with the Inspector Gadget lensing simulation pipeline \citep{InspectorGadget1, InspectorGadget2} on the New York Blue supercomputer, which is part of the New York Center for Computational Sciences at Brookhaven National Laboratory/Stony Brook University. In this section we describe the simulations and the cosmological models we chose to study.

\subsection{N-body Simulations}

The N-body simulations are the same as those used in \citet{Yang:2011zz}, \citet{Minkowski} and \citet{Xiuyuan2}, and consist of a series of 80 CDM $N$-body simulations with $512^3$ particles each and a box size of $240h^{-1}$~Mpc. They were run with a modified version of the public N-body code Gadget-2 \citep{Springel:2005mi}. The linear matter power spectrum, which serves as input for the initial conditions generator N-GenIC associated with Gadget-2, was created with CAMB \citep{Lewis:1999bs} for $z=0$, and scaled to the starting redshift of our simulations at $z=100$ according to the linear growth factor.

The N-body simulations cover different cosmological models produced in multiple runs with different random initial conditions. A total of 50 of the runs is available in the fiducial cosmology, with parameters chosen to be \{$\Omega_m=0.26$,$\Omega_\Lambda=0.74$, $w=-1.0$, $n_s=0.96$, $\sigma_8=0.798$, $H_0=0.72\}$.  
These 50 runs all used the same input power spectrum, but each one is a different and strictly independent realization. 
This yields a statistically robust set of simulations. 
In each of the other six cosmological models one parameter was varied at a time, keeping the others fixed, with the following values: $\Omega_m=\{0.23, 0.29\}$ (while $\Omega_\Lambda=\{0.77, 0.71\}$ such that the universe stays spatially flat), $w=\{-0.8, -1.2\}$, and $\sigma_8=\{0.75, 0.85\}$. 
For each of these six non-fiducial cosmological models 5 simulations are available, where each simulation used a different realization of the initial conditions. 
Table \ref{tab:Cosmologies} lists all the cosmological models with their parameters and number of N-body simulations.

\begin{table}
\begin{centering}
\caption[]{\textit{Parameters varied in each cosmological model and weak lensing map set. }}
\begin{tabular}{|l|c|c|c|c|c|} 
\hline
WL Map Set & $\sigma_8$ & $w$ & $\Omega_m$ & $\Omega_\Lambda$ & \# of \\
Identifier & & & & & sims \\
\hline
Primary & 0.798 & -1.0 & 0.26 & 0.74 &45\\
Auxiliary & 0.798 & -1.0 & 0.26 & 0.74 & 5\\
Om23 & 0.798 & -1.0 & 0.23 & 0.77 & 5\\
Om29 & 0.798 & -1.0 & 0.29 & 0.71& 5\\
w12 & 0.798 & -1.2 & 0.26 & 0.74 & 5\\
w08 & 0.798 & -0.8 & 0.26 & 0.74 & 5\\
si75 & 0.750 & -1.0 & 0.26 & 0.74 & 5\\
si85 & 0.850 & -1.0 & 0.26 & 0.74 & 5\\
\hline
\end{tabular}\label{tab:Cosmologies}
\end{centering}
\end{table}

The shear and convergence maps, described in more detail in the next subsection, were generated by mixing simulations with different random initial conditions, and by randomly rotating and shifting the simulation data cubes. For the maps in each non-fiducial cosmology a mixture of all five independent N-body runs was used. In the fiducial cosmology, two completely independent sets of maps are available. One of these sets, called hereafter the ``auxiliary'' map set, was created from the five independent N-body runs with the same five quasi-identical\footnote{By ``quasi-identical'', we mean that the random number seeds to create the initial particle distributions from the power spectra were kept the same across all cosmological models, but the normalization of the power spectra themselves were adjusted such as to yield the desired $\sigma_8$ today in every cosmology. These adjustments are necessary due to the difference in growth factors between the models.} initial conditions as in the non-fiducial cosmologies. The second map set was created by mixing lens planes from the remaining larger ensemble of 45 independent N-body runs, and will be referred to as the \lq\lq primary\rq\rq\ map set. This is also reflected in Table \ref{tab:Cosmologies}.

\subsection{Weak Lensing Maps}

Our pipeline uses a standard two-dimensional ray-tracing algorithm, as described in~\citet{Hamana:2001vz}, to create the weak lensing maps from the N-body simulations. 
Earlier work using similar algorithms includes ~\citet{Schneider:1992},  \citet{Wambsgaans:1998} and \citet{Jain:1999ir}. 
We refer the reader to \citet{Kratochvil:2009wh, Minkowski, Yang:2011zz} for the full description of our methodology and verification of the accuracy of the simulations used.  

The large-scale structure from the N-body simulations was output as particle positions in boxes at different redshifts, starting at redshift $z=2$.  
The particles were then projected perpendicularly onto planes spaced $80h^{-1}$Mpc apart in a direction parallel to the central line of sight of the map. We used the triangular shaped cloud (TSC) scheme \citep{Hockney-Eastwood} to place the particles on a grid on these two-dimensional density planes; the particle surface density was then converted into the gravitational potential via the Poisson equation. The algorithm then followed light rays from the observer back in cosmic time. The deflection angle, as well as the weak lensing convergence and shear were calculated at each plane for each light ray. These depend on the first and second derivatives of the potential, respectively. Between the planes, the light rays traveled in straight lines.

Shear and convergence maps, spanning 12 square degrees, were created for $2048\times2048$ light rays. For simplicity, we created maps assuming the source galaxies to be at three fixed redshifts, $z_s=1, 1.5, 2$.
Each cosmological model is represented by 500 such 12-square-degree maps for convergence and shear parameters for each of the three source galaxy redshifts.

%% file: galaxyModel.tex
\section{Source Galaxies}
\label{sec:gal_mod}

In this section we describe how we characterize the source galaxies which we use to trace the shear field. 
We wish to make our prediction for shear peak counts as realistic as possible, and for that it is essential that we make our source galaxies as realistic as possible. 
The steps we take to create the ensemble of source galaxies can be summarized as follows:
\begin{itemize}
\item Assign a spatial position for the galaxy. 
\item Assign a redshift for the galaxy.
  Based on redshift, assign the galaxy a magnitude, size and intrinsic shape. 
\item Add reduced shear to galaxy. 
  Re-calculate size and magnitude. 
\item Add reduced shear error to galaxy. 
\end{itemize}

\subsection{Intrinsic Properties}
As described in Section~\ref{sec:sims}, we have 500 realisations of maps for each of 7 different cosmological models.  
Each set consists of ray-traced maps of the lensing parameters $\gamma_1$, $\gamma_2$ and $\kappa$ in three redshift bins, at $z$=1.0, 1.5 and 2.0. 
We consider each of the 500 realisations of one cosmological model to be independent observations of the sky, and for each map we generate an independent ensemble of galaxies to use as tracers of the shear field in three dimensions. 
The same source galaxies are used for all 7 cosmologies, so we are effectively observing the same \lq\lq sky\rq\rq\ with all cosmological models.

We scatter the galaxies randomly across the field, ensuring that we have an average  galaxy density of 30 galaxies arcmin$^{-2}$, which is roughly the expected galaxy density usable for weak lensing analyses for an LSST ten-year survey in r-band~\citep{lsst-wl}. 
At this first step, we have already limited how realistic we can make this study: in randomly positioning the source galaxies, we do not take into account the shifts in their apparent positions due to lensing and that the source galaxy positions are in reality correlated with dark matter halos in the simulation. 
We decided to neglect these effects because it lets us shoot light-rays backwards in time through the N-body simulation indiscriminately, as opposed to having to determine which light ray hits a fixed galaxy position. 
Matching galaxy density with input shear maps is very difficult; see for example~\citet{behroozi}.

One consequence of these simplifications is that we neglect the magnification bias present in lensing (\citet{magnification bias papers}).  
The magnification bias arises from two competing effects: i) high-shear regions magnify galaxies, thus making fainter galaxies visible in a flux-limited survey and adding source galaxies in those regions of the sky, ii) the magnification also spreads apart the apparent positions of the source galaxies, thus diluting the number of galaxies in these high-shear regions. 

We anticipate that this variation in density will have a small impact on lensing peak counts, or at least on the cosmological constraints coming from lensing peak counts. 
This is because the constraints have been shown to be dominated by the numerous low peaks \citep{Kratochvil:2009wh, Dietrich:2009jq}, which are to be found in low shear regions, while the magnification bias is most noticeable in regions of high shear. 
For the high significance lensing peaks, the primary effect of magnification on galaxies will be a shift in the apparent position of galaxies which can also shift the position of a peak (particularly if it is not the central peak of a cluster). 
Since peak counts (measured using one smoothing scale as done in this paper) do not measure angular correlations, a shift in position will not affect the results. 
For the central peak of a cluster, however, the dilution of source galaxies associated with magnification will mostly cause an apparent broadening of the peak, which will make the peak appear larger than in our simplification. 
We do not expect this to be a significant effect for the cosmological constraints, because high central peaks are by far outnumbered by the others, but the importance of this effect should be studied in future work.

\begin{figure}[htb]
\includegraphics[width=3.5in]{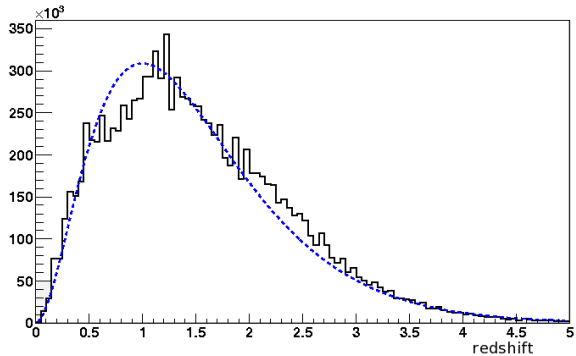}
\caption{Number of galaxies as a function of redshift for the source galaxies. The blue dashed line shows a fit to the simple model  $n(z)\propto z^2 e^{-2z}$. }
\label{fig:gals-z}
\end{figure}

Next we assign each galaxy a redshift, size and magnitude, taken from a distribution obtained from the input catalogues of the LSST Image Simulator.
Galaxies in these catalogues have properties based on those produced by the Millennium dark matter simulations. 
The galaxy catalogue is complete out to an $r$ magnitude of 28, which is approximately one magnitude deeper than the expected depth of the full LSST ten-year survey. 
These quantities have been anchored to observations from a compilation of deep survey data\footnote{http://astro.dur.ac.uk/\textasciitilde nm/pubhtml/counts/counts.html}, the DEEP2 survey~\citep{coil04}, and data from the publicly available Hubble Deep Field catalogues\footnote{http://www.stsci.edu/ftp/science/hdf/archive/v2.html}. 
A redshift is assigned at random to the galaxies, shown in Figure~\ref{fig:gals-z}, where the dashed line represents a simple model of the form $n(z)\propto z^2 e^{-2z}$, as described in~\citet{wittman} and previously found to be a good fit to DEEP2 survey data~\citep{coil04}. 
A redshift-dependent size and magnitude is assigned for each galaxy from the simulated input catalogues, where we define size as the product of the RMS of the semi-major and semi-minor axes of the galaxy.

An intrinsic ellipticity is then assigned to each galaxy. 
We base the assigned ellipticity on measurements made using COSMOS data~\citep{joachimi, leauthaud}, where the intrinsic galaxy shape noise was found empirically to be 0.23 per reduced shear component. 
A small dependence on galaxy magnitude was identified in the COSMOS data, but no significant dependence on size or redshift was found. 
We next assign a reduced shear to the galaxy, taken from the input shear and convergence maps described in Section~\ref{sec:sims} and extrapolated to the galaxy position in RA, DEC and redshift (where we linearly extrapolate the shear and convergence parameters between the three redshift planes at z=[1.0,1.5,2.0]). 
The galaxy size, magnitude and ellipticity are re-calculated to account for the effect of shearing and magnification.

\subsection{Measurement Error}

To assign a measurement error for the galaxy shape is somewhat complicated. 
Our aim is to obtain an error consistent with a measurement of galaxy shapes from a ten-year stack of LSST observations, using only observations made with good seeing in the $r$-band. 
If we require the median seeing to be 0.66$\arcsec$, which is an acceptable quality for weak lensing analyses~\citep{lsst-wl}, we expect a ten-year stack to consist of approx.\ 100 observations.

We use the LSST Photon Simulator (PhoSim) to simulate a ten-year stack of LSST observations of an area on the sky the size of a LSST chip ($\frac{1}{16}$ the area of an LSST CCD, with approx. 13.6 arcmin$^2$ field of view), at different positions on the LSST focal plane in order to sample the PSF as it varies across the focal plane. 
PhoSim is a high-fidelity, ray-traced end-to-end simulator of the LSST system. 
A detailed description of the system can be found in~\citet{peterson}, \citet{imsim} and \citet{chihway-atmos}.  
Recent work by~\citet{chihway-wl} has studied in depth the impact on galaxy shape measurement made by the distortions introduced by the atmosphere and the LSST telescope itself. 
 We wish to isolate the impacts of measurement and algorithmic effects, and to evaluate the impact of these errors separately from the error due to galaxy intrinsic shape noise. 

Since we have already accounted for shape noise in a previous step in the pipeline, we use in these simulations an input catalogue consisting of circular galaxies with a Gaussian profile to remove any effects of shape noise. 
The magnitudes, redshifts, SNRs and spectral energy distributions of these Gaussian galaxies are the same as the fully realistic galaxy distribution of the ImSim input catalogues. 
The advantage to this approach is that we can easily evaluate the measurement error without attempting to remove galaxy shape noise. 
We are not performing a redshift-dependent measurement, and we ignore for this work the potentially substantial errors in redshift measurement.

We take 500 values of [$\gamma_1, \gamma_2, \kappa$] at random from one of the simulated shear maps described in Section~\ref{sec:sims}, and produce 500 different sheared ImSim input catalogues by applying a single reduced shear value [$g_{1,in}, g_{2,in}$] to all galaxies in an existing catalogue. 
For each of these sheared input catalogues, we produce 100 simulated images of the same area of sky, each time with different atmospheric conditions specified by the LSST Operations Simulator~\citep{opsim} selected such that the median seeing is 0.66$\arcsec$. 
We process the resulting images using the SourceExtractor object-detection package~\citep{sEx}. 
For each exposure, we use the stars in the field to reconstruct the PSF, which is  interpolated to the galaxy locations using a third order polynomial interpolation function. 
The measured galaxy shapes are corrected for distortions due to the PSF using  the popular KSB~\citep{ksb} algorithm implemented in the IMCAT\footnote{http://www.ifa.hawaii.edu/{\textasciitilde}kaiser/imcat/} pipeline. 
We use the KSB algorithm because it is well known in the community, and its strengths and weaknesses are well understood. 
For example, it is known that the process that converts ellipticity to reduced shear should be calibrated using simulations. 
We apply a ``perfect'' calibration, by applying a calibration factor that shifts the mean measured  reduced shear in each of our simulated exposures to the input  reduced shear value. 
The measured, PSF-corrected, calibrated shape for each galaxy is then averaged over the 100 atmospheric realisations, giving us an estimate of the stacked galaxy shape measurement. 
More sophisticated algorithms are expected to give a smaller uncertainty on galaxy shape measurement (see ~\citet{great10} for a summary of the performance of many current shape measurement methods). 
Despite applying a perfect calibration, for this reason we consider the uncertainty obtained from our KSB implementation to be conservative for LSST.

\begin{figure}[htb]
\includegraphics[width=3.5in]{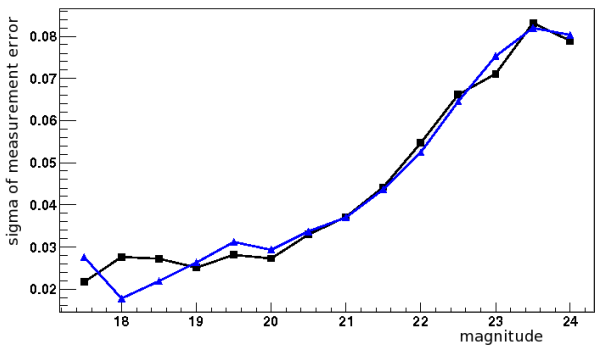}
\caption{Standard deviation of the errors for measured cosmic shear parameters $g_{1,in}$ and $g_{2,in}$, for circular galaxies of different $r$ magnitudes. The standard deviation is taken from fitting a Gaussian to the distribution of galaxy shape measurement errors for each magnitude bin. Black squares represent g1, blue triangles g2. }
\label{fig:gals-sigma-mag}
\end{figure}

We compare the measured galaxy shapes to the input reduced shear values, and the difference between input and output gives the uncertainty on the reduced shear measurement. 
There is a dependence of the measurement uncertainty with magnitude, with fainter galaxies having larger uncertainties, as shown in Figure~\ref{fig:gals-sigma-mag}. 
This is in accordance with the dependence of reduced shear measurement uncertainty with object magnitude found in data from the COSMOS survey~\citep{leauthaud}. 
We account for this dependence as we assign measurement errors drawn from this distribution, which are added to the galaxy shape noise. 
Since we assign the noise to galaxies randomly, we do not consider any spatial correlations the noise may have across the field. 
\citet{chihway-wl} has shown that, for a ten-year stack of LSST images, the spatial correlation of measurement error (including an imperfectly modeled PSF) is at a level comparable to the statistical error on the weak lensing correlation function, around $10^{-7}$. 
We therefore consider that the spatial correlations of measurement error will be similarly small for aperture mass statistics, and neglect it in this work. 
However, future work is planned to specifically quantify the impact of correlated error on peak counts.

\begin{figure}[htb]
\includegraphics[width=3.5in]{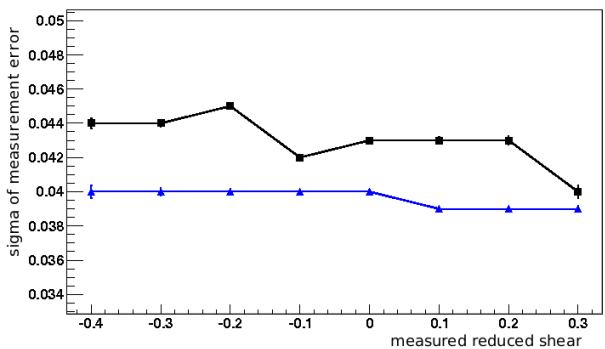}
\caption{Standard deviation of the errors for input reduced shear parameters $g_{1,in}$ and $g_{2,in}$, for elliptical galaxies of different measured reduced shear $g_{1,meas}$ and $g_{2,meas}$. The standard deviation is taken from fitting a Gaussian to the distribution of galaxy shape measurement errors for each measured $g$ bin. Black squares represent $g_1$, blue triangles $g_2$. }
\label{fig:gals-sigma-rg}
\end{figure}

We also wish to investigate the dependence of measurement error with galaxy shape. 
To do this, we therefore made a set of simulations identical to those described above, but using elliptical galaxies. 
However, we are unable to separate the intrinsic shape noise from the measurement error in these simulations, so we are limited to examining the dependence of the total error on the measured reduced shear with the measured galaxy shape. 
Figure~\ref{fig:gals-sigma-rg} shows the uncertainty distribution for input reduced shear values [$g_{1,in},g_{2,in}$] for different values of measured galaxy shape [$g_{1,meas}, g_{2,meas}$]. 
There is no dependence with galaxy shape, and the distribution is remarkably flat. 
We do see a significant dependence of the uncertainty on [$g_{1,in}, g_{2,in}$] with the raw ellipticity measurement of the galaxy, but the process of PSF deconvolution using KSB, and the calibration procedure removes this dependence. 
We therefore do not apply a shape-dependent measurement error.

\begin{figure}[htb]
\includegraphics[width=3.5in]{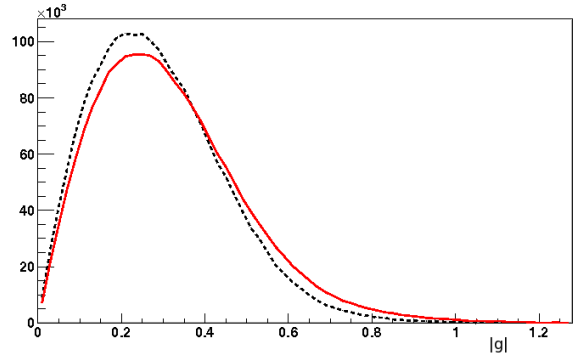}
\caption{Distribution of galaxy shapes $|g|$, for intrinsic ellipticity only (black dotted line) and including measurement error (red solid line).}
\label{fig:rg-int-meas}
\end{figure}

Figure~\ref{fig:rg-int-meas} shows the distribution of the assigned values of the reduced shear $|$g$|$ for all galaxies in our sample, comparing the intrinsic ellipticity alone to the combination of intrinsic ellipticity and measurement error. 
The measurement error has a much smaller contribution to the total galaxy shape error than the intrinsic ellipticity, but is not negligible. 
We shall see in the next sections what impact this has on the peak counts and cosmological constraints. 

%% file: peakcounts.tex
\section{Peak Counts}
\label{sec:peakcounts}

\begin{figure}[htb]
\includegraphics[width=3.5in]{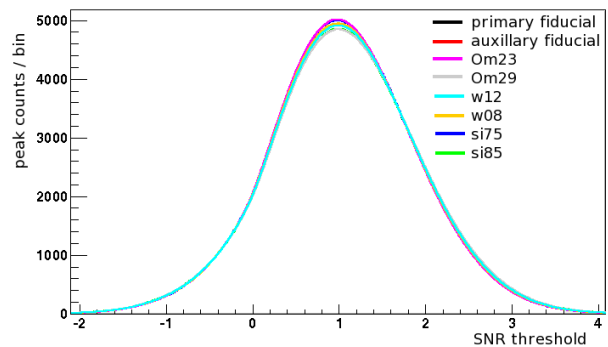}
\caption{Peak counts above SNR threshold for different cosmological models, for the aperture mass calculated using realistic galaxies with both intrinsic shape noise and measurement error. See Table~\ref{tab:Cosmologies} for details of cosmological model parameters.  }
\label{fig:counts-rgErr}
\end{figure}

We consider these mock galaxy shape measurements to be a representative sample of an LSST ten-year survey. 
We use these simulated datasets to perform the aperture mass calculation given in Section~\ref{sec:formalism} using the GPU implementation described in ~\citet{gpu}, and obtain peak counts for the seven cosmological models described in Table~\ref{tab:Cosmologies}. 
The aperture mass is calculated using source galaxies with shape noise alone, and using source galaxies with both shape noise and measurement error.  
Figure  ~\ref{fig:counts-rgErr} shows the distributions of peak counts for the different cosmological models, where each model is sampled by the same galaxies (including  shape noise and measurement errors) scaled to the full-sky LSST survey size. 

\begin{figure}[htb]
\includegraphics[width=3.5in]{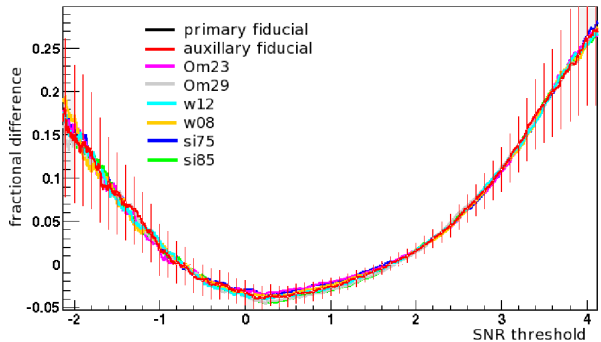}
\caption{Fractional difference between peak counts with shape noise alone, and peak counts with shape noise and measurement error, given as the fractional difference from shape noise only. Solid curves are the average over the 500 different maps; error bars are the standard deviation of the 500 maps, shown for the auxiliary fiducial model to indicate the level of statistical error. See Table~\ref{tab:Cosmologies} for details of cosmological model parameters.  }
\label{fig:counts-rgInt-rgErr}
\end{figure}

To evaluate the impact of measurement error, compared to shape noise alone, we calculate the difference between the peak counts for the two cases. 
This is shown in Figure~\ref{fig:counts-rgInt-rgErr}, where we plot the fractional difference between the peak counts for intrinsic shape noise alone, compared to shape noise and measurement error. 
The difference is largest at very low and high peak significance, where it reaches up to 25\%. 
As we might expect, the difference is identical for all cosmological simulations, showing that measurement error should not bias constraints on cosmological parameters in favour of one model over another. 

\begin{figure}[htb]
\includegraphics[width=3.5in]{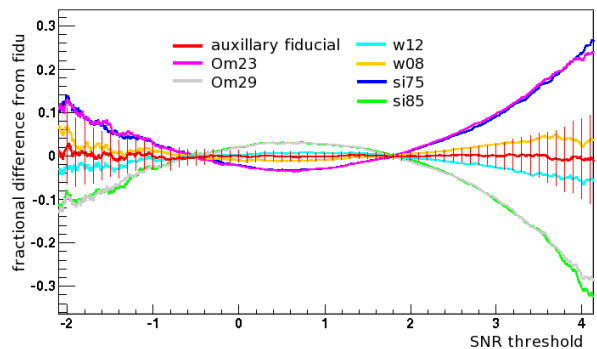}
\caption{Difference between peak counts in fiducial cosmology, and peak counts in other cosmologies for aperture mass calculated using realistic galaxies with both shape noise and measurement error, given as the fractional difference from fiducial cosmology counts above SNR threshold.  Solid curves show the mean difference for the 500 maps used in the measurement; error bars are the standard deviation, shown for the auxiliary fiducial model to indicate the level of statistical error. See Table~\ref{tab:Cosmologies} for details of cosmological model parameters. }
\label{fig:counts-rgErr-diff}
\end{figure}

It is hard to distinguish between the different cosmological models by eye in Figure ~\ref{fig:counts-rgErr}, but if we plot the difference of the peak counts from the fiducial cosmology, as in Figure ~\ref{fig:counts-rgErr-diff}, we obtain a clearer view of the characteristics of each cosmology. 
The red curve in these figures represents the peak counts obtained from the auxiliary WL map set of the fiducial model, and acts as a control test to be compared with the primary map set of the fiducial model belonging to the same cosmology. 
In Figure~\ref{fig:counts-rgErr-diff} it is clear that the difference between the primary and auxiliary map sets of the fiducial model are consistent within the statistical error (shown by the error bars on the auxiliary fiducial model). 
Several of the cosmological models have very similar peak count distributions. In accordance with expectations of the ($\sigma_8, \Omega_m)$-degeneracy, the models Om23 and si75, and Om29 and si85, have very similar peak count profiles, which will result in our predicted cosmological constraints exhibiting a degeneracy in the corresponding direction in the parameter space.

\begin{figure}[htb]
\includegraphics[width=3.5in]{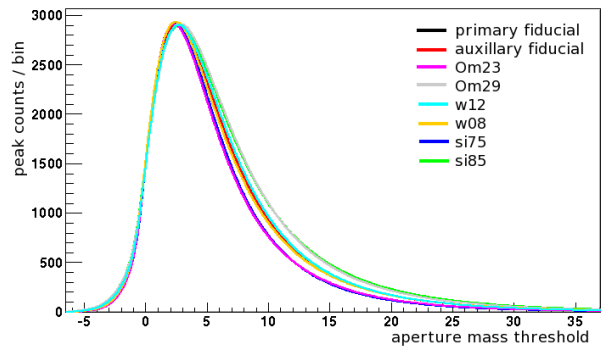}
\caption{Peak counts above SNR threshold for different cosmological models for the ``perfect'' case where the aperture mass is calculated directly from the reduced shear maps. See Table~\ref{tab:Cosmologies} for details of cosmological model parameters.  }
\label{fig:counts-rg}
\end{figure}

\begin{figure}[htb]
\includegraphics[width=3.5in]{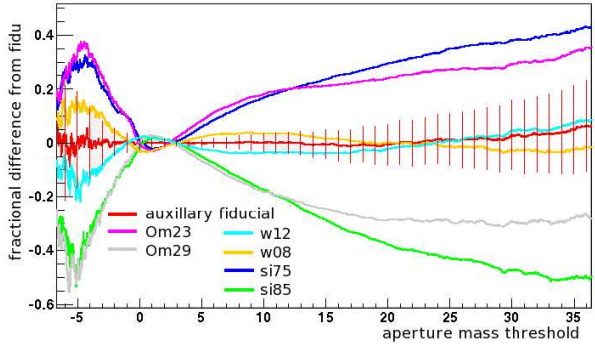}
\caption{Difference between peak counts in fiducial cosmology, and peak counts in other cosmologies for the ``perfect'' case where the aperture mass is calculated directly from the reduced shear maps, given in \%\ difference from fiducial cosmology counts above SNR threshold.   Solid curves show the mean difference for the 500 maps used in the measurement; error bars are the standard deviation, shown for the auxiliary fiducial model to indicate the level of statistical error. See Table~\ref{tab:Cosmologies} for details of cosmological model parameters. }
\label{fig:counts-rg-diff}
\end{figure}

We can also compare these peak counts to the counts obtained from calculating the aperture mass directly from the maps of reduced shear, without sampling the maps with source galaxies. 
This is the ``perfect'' case where we have perfect knowledge of the shear, and no noise is introduced by galaxy shape noise or measurement errors. 
It is therefore an impossible ideal, but serves as a useful comparison to examine how real measurements are affected by error. 
Since there is no noise in this measurement, constructing a SNR map is meaningless, and instead we count peaks in the map of aperture mass. 
The two quantities can be related by
\begin{equation*}
M_{ap}(\theta_0) = \mathrm{SNR}(\theta_0)  \hat{\sigma}_{M_{ap}}
\end{equation*}
where $M_{ap}$, SNR and $\hat{\sigma}_{M_ap}$ are defined in Equations~\ref{eqn:map}, \ref{eqn:snr} and \ref{eqn:sigmap} respectively. 
Figures~\ref{fig:counts-rg} and ~\ref{fig:counts-rg-diff} show the peak counts above aperture mass thresholds for different cosmologies, and the difference of the seven cosmological models compared to the primary map set of the fiducial model, respectively. 

The addition of shape noise and measurement error has a significant impact on the shape of the peak counts, visible in a comparison of Figures~\ref{fig:counts-rgErr-diff} and~\ref{fig:counts-rg-diff}. 
For the realistic case with errors included, the overall shape of the peak counts will be valuable in constraining cosmological parameters, since the deviation from the fiducial cosmology is visible at all SNR levels. 
The addition of noise has impacted both the significance of the peaks, as one would expect, and also the shape of the peak counts of the different cosmologies.

\subsection{The Significance of Low Peaks}

In this section we discuss further the significance of low peaks. 
Both \citet{Dietrich:2009jq} and \citet{Kratochvil:2009wh} discovered that low peaks contribute the most to the cosmological constraints, which was later confirmed and elaborated upon by others \citet{Yang:2011zz, Minkowski, Marian:2011rg, Maturi:2011am, Xiuyuan2}. 
What has gone mostly unnoticed is that the different papers have completely different definitions of the word \lq low\rq\ in this context, and different papers actually refer to completely disjoint peak ranges. 
\citet{Kratochvil:2009wh} and the group's follow-up works \citep{Yang:2011zz, Xiuyuan2}, subsequently referred to as Group A, define low peaks as having an SNR between 0--2 or 1--3.5 $\sigma$, depending on the publication.  
\citet{Dietrich:2009jq} define low peaks as lying in the range 3.25--4.5$\sigma$, such that their entire range is higher than most of the previous group's papers. 
The hierarchical peak finding algorithm of \citet{Marian:2011rg} is overwhelmed by the number of peaks below SNR$\sim3\sigma$ and breaks down, so these authors also restrict themselves to a range above SNR$\sim3\sigma$ in their peak detection, while \citet{Maturi:2011am} conclude that to constrain the parameter $f_{NL}$ of primordial non-Gaussianity, only peaks with SNR$>2\sigma$ are useful. 
We refer to this second group of three independent collaborations as Group B.

What has made a direct comparison of the works of these papers impossible is that Group A used peak counts in maps of convergence, while Group B used peak counts in shear maps and a somewhat more realistic simulation of galaxy shape noise. 
For the first time in this paper, we use reduced shear maps and include realistic LSST measurement errors for the same set of simulations used by Group A, which allows a direct comparison of the works. 
However, it should be noted that our comparison is not complete, since we use only one smoothing scale and a different aperture mass filter compared to the work in Group A. 

\begin{figure}[htb]
\includegraphics[width=3.5in]{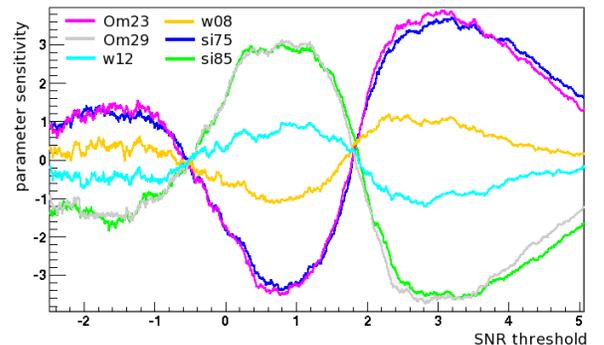}
\caption{The $\Delta\chi^2$ (parameter sensitivity) from peak counts for our non-fiducial cosmological models, shown as a function of peak height in units of signal-to-noise ratio. Neglecting correlations between the different heights, positive peaks with $SNR<2\sigma$ carry approximately 1/3 of the information content, those with SNR$>3\sigma$ carry one half, while peaks with SNR$>3.5\sigma$ carry about another third. See Table~\ref{tab:Cosmologies} for details of cosmological model parameters.}
\label{fig:counts-chi2-rg-diff}
\end{figure}

We plot the $\Delta\chi^2$ between different cosmological models coming from the different SNR ranges in Figure \ref{fig:counts-chi2-rg-diff}. 
Neglecting correlations between individual SNR ranges and simply interpreting the area under the curves as the strength of distinction between the cosmological models, we conclude that peaks with SNR$\sim0-2\sigma$ carry approximately 1/3 of all the information in the peak counts, peaks with SNR$>3\sigma$ approximately half, and peaks with SNR$>3.5\sigma$ also about 1/3. 
We can compare this result to \citet{Kratochvil:2009wh}, which found according to the third panel of their Figure 5 that low peaks (by our definition of \lq low\rq ) were somewhat more important for cosmology with convergence maps.  
However, we use a filter for aperture mass that emphasizes an NFW profile and so may de-emphasize smaller peaks, which would explain the discrepancy in our results.\footnote{Also see \citet{Pires_PC:2012} which claims that convergence can contain complementary information to shear if one manages to extract it observationally.} 

In the literature, there have been claims that low (0-3$\sigma$) peaks do not carry any useful cosmological information both due to galaxy shape noise dominating the peak counts, and due to the unknown influence of systematic errors in this range. 
\citet{Yang:2011zz} has shown that the first of these issues is a misconception, and that real cosmological structure contributes significantly to peak counts at low SNR. 
We have shown here that, even in the presence of systematic errors from a realistic analysis pipeline, we can still extract that information.

Interestingly, an important result from \citet{Xiuyuan2} is that peaks with SNR$\sim1-3.5\sigma$ are largely unbiased by baryon effects and therefore lend themselves particularly well for cosmological parameter estimations at the sub-percent accuracy level to which LSST aspires. Their result is valid within the restrictions of their study of varying the concentration parameter within NFW halos. To obtain certainty that this is universally the case for all possible contributions of baryonic physics, full hydrodynamic simulations with different baryon prescriptions need to be run. We are in the process of investigating this for a future publication. 

We conclude that there is substantial information content in low WL peaks, with \lq low\rq\ defined as SNR$\sim0-3\sigma$. 
Low peak counts should not be dismissed as purely due to shape noise, especially since these low peaks have been shown to be less sensitive to uncertain baryonic physics in simulations.

%% file: analysis.tex
\section{Analysis}
\label{sec:analysis}

We present in this section the methodology for extracting cosmological information from the peak counts and power spectra we have obtained from aperture mass calculations. 

\subsection{Statistical Descriptors} 

We generically refer to the different statistics one can obtain from a 2D WL map---e.g. power spectrum, peak counts, etc.---as statistical \lq\lq descriptors\rq\rq, and denote them by $\mathbf{N}$. 
We can also combine them into a single vector, $N_i$, where $i$ indexes the peak height or the multipole for the power spectrum. 
Combining the data from several source redshifts makes the descriptor vector longer, but it is treated in the same way. 
We divide the range of peak height into 30 threshold bins.  Similarly, we divide the power spectrum into 30 scale bins.

To constrain cosmology, we are interested in the true ensemble average\footnote{Averaged
over all possible universes with the same cosmological parameter values.}
(denoted henceforth by brackets $\langle\ \rangle$) and covariance of
these descriptors as a function of cosmological parameters
($\mathbf{p}=\{\Omega_m, w, \sigma_8\}$).  
These of course are not available to us, but can be estimated from the simulations.  
Averaging over the pseudo-independent map realizations within a given cosmology, we can estimate the ensemble average by 
\BE\label{SM}
\langle{N}_i(\mathbf{p})\rangle\approx\overline{N}_i(\mathbf{p})\equiv\frac{1}{R}\sum_{r=1}^R
N_i(r, \mathbf{p}),
\EE
where $N_i(\mathbf{p})$ is the descriptor vector for one set of cosmological parameters $\mathbf{p}$, $N_i(r, \mathbf{p})$ is the descriptor vector for a single realization and $r$ runs over our $R=500$ map realizations.  
We call
this estimate the \textit{simulation mean}.  
It differs from the true ensemble average both because of the limited number of realizations and also because of the limitations inherent in our simulations.  
In the absence of a fitting formula for the peak counts in the non-Gaussian case  \citep[analogous to the power spectrum formula from][]{Smith} the simulation mean serves as our proxy for theoretically predicted peak counts.\footnote{While there are theoretical predictions for peak counts in Gaussian and non-Gaussian cases, the non-Gaussian predictions are not accurate enough to be useful for this purpose.}

Because of the computational expense of producing cosmological simulations, we can only produce a limited number at the selected cosmologies given in Table~\ref{tab:Cosmologies}.  
We have calculated the simulation mean at these points and must extrapolate to other cosmologies not explicitly simulated. 
Using finite differences between the simulated cosmologies, we construct a first-order Taylor expansion around our fiducial cosmology to estimate $\overline{N}_i(\mathbf{p})$ for other cosmologies: 
\BE\label{Taylor}
\overline{N}_i(\mathbf{p})\approx\overline{N}_i(\mathbf{p_0})+\sum_\alpha \frac{\overline{N}_i(\mathbf{p}^{(\alpha)})-\overline{N}_i(\mathbf{p_0})}{p^{(\alpha)}_\alpha-p_{\mathbf{0}\alpha}}\cdot(p_\alpha-p_{\mathbf{0}\alpha}),
\EE
Here, index $\alpha=1,2,3$ refers to an individual parameter, such as $\Omega_m, w$, or $\sigma_8$. 
 $\mathbf{p}^{(\alpha)}$ denotes the cosmological parameter vector of a simulated non-fiducial cosmology (where only the parameter $p_{\alpha}$ has been varied), and $\mathbf{p_0}$ denotes the parameter vector for the fiducial cosmology.

The fraction in Eq.~(\ref{Taylor}) is the finite difference derivative. 
If the non-fiducial cosmology is chosen such that $p^{(\alpha)}_\alpha-p_{\mathbf{0}\alpha}$ is positive, we call it a ``forward derivative'', if it is negative, we call it a ``backward
derivative''. 
We compare the parameter constraints calculated from each derivative  to assess the robustness of our results. 

Similarly to the simulation mean, we estimate the covariance of the statistical descriptors from the simulations, ${\rm Cov}( N_i, N_j) \approx C_{ij}$, where
\BE
\label{covariance matrix}
C_{ij}(\mathbf{p})\equiv\frac{1}{R-1}\sum_{r=1}^R [N_i(r,\mathbf{p})-\overline{N}_i(\mathbf{p})][N_j(r,\mathbf{p})-\overline{N}_j(\mathbf{p})].
\EE
This covariance matrix contains contributions both from the sample
variance of the true aperture mass signal and from the noise contributions.  
When the size of this covariance matrix is large, inaccuracies in its estimate can become challenging, as we explore further below.

\subsection{Monte Carlo Probability Contours}\label{sec:Monte Carlo}

Each of our WL maps spans a 12 deg$^2$ field of view, yet we wish to obtain parameter contours for the full 20,000 deg$^2$ LSST survey volume. 
We therefore employ bootstrapping to generate approximations to full-sky maps. 
In this procedure, we draw a map $20,000/12\approx1667$ times from our 500 aperture mass maps, with replacement. 
The resulting 20,000 deg$^2$ map is not a true composite: we do not place the drawn maps edge to edge, but rather compute the descriptor values for each patch individually and then average over them to get their values for the full-sky map. 
Details of this method, as well as its advantages for parameter estimation, will be discussed in this context in \citet{InspectorGadget2}. 
We create 10,000 such full-sky maps to obtain smooth parameter contours in our Monte Carlo procedure. 

To estimate the cosmological parameter error contour from a set of WL maps from one cosmology, we use $\chi^2$-minimization to fit for the best-fit cosmological parameters for each of the above bootstrapped full-sky maps. 
Thus our whole set of maps provides an ensemble of Monte Carlo realizations, and the distribution of those best-fit points can be used to draw probability contours at desired confidence levels.

For realizations drawn from the fiducial cosmology $\mathbf{p_0}$, $\chi^2$ is
\BE 
\label{chi2}
\chi^2(r, \mathbf{p})\equiv\sum_{i,j}\, \Delta N_i(r, \mathbf{p})\,[{\rm Cov}^{-1}\mathbf{(p_0)}]_{ij}\, \Delta N_j(r,\mathbf{p})
\EE
where
\BE
\Delta N_i(r,\mathbf{p})\equiv N_i(r,\mathbf{p_0})- \langle{N}_i(\mathbf{p})\rangle.
\EE

For each Monte Carlo realization, we minimize $\chi^2$ with respect to $\mathbf{p}$ using a simulated annealing algorithm.  
As outlined in \citet{InspectorGadget2}, the covariance matrix is computed from the auxiliary map set and inverted with singular value decomposition, discarding any problematic eigenvectors. 
The simulation mean for the fiducial cosmology is computed from the primary map set, whereas the finite difference derivatives are computed from the auxiliary map set. 
The maps for which best-fit parameters are computed come from the primary map set.

%% file: cosmology.tex
\section{Cosmological Constraints}
\label{sec:cosmo}
In this section we present the results obtained by applying the methods described in Section~\ref{sec:analysis} to the peak counts described in Section~\ref{sec:peakcounts} in order to constrain the cosmological parameters $\Omega_m$, $\sigma_8$ and $w$. 
We also calculate the power spectrum of the aperture mass maps, and use that to predict constraints on cosmological parameters. 
Previous work~\citep{Yang:2011zz} has shown that peak counts in convergence maps contain additional information beyond the power spectrum, and can tighten constraints on cosmological parameters, as well as break degeneracies in constraints from the power spectrum alone. 
We compare here the constraints on cosmological parameters obtained from the power spectrum of the aperture mass maps, traced by galaxies with shape noise included, and the constraints from peak counts, traced by the same galaxies. 

To determine the sensitivity of peak counts to cosmological parameters, one can use the backward, forward, or symmetric derivative in the Taylor expansion in Eq.~(\ref{Taylor}). 
It is important to check that all derivatives give the same contours, and as expected we find very small shifts in the contour size and shape between the different derivative methods consistent with the statistical limitations of the cosmological simulations used. 
The contours we present in this section are 68\%\ error contours, corresponding to data obtained by a full LSST ten-year survey using only good quality $r$-band data.

\begin{figure*}[htb]
\includegraphics[width=7.1in]{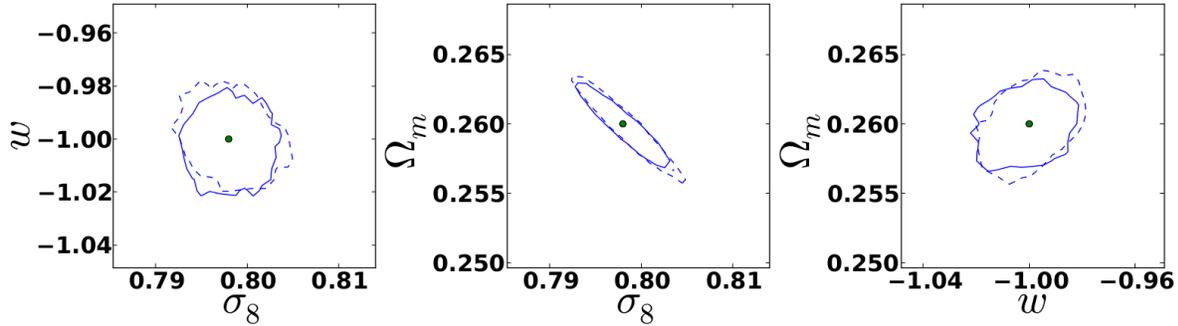}
\caption{68\%\ error contours on the cosmological parameters $\Omega_m$, $\sigma_8$ and $w$ for peak counts, using the backward derivative in the Taylor expansion. The dashed curves show the constraints for measurements including shape noise only, and the solid curves for both shape noise and measurement error. }
\label{fig:contours-back}
\end{figure*}

\begin{figure*}[htb]
\includegraphics[width=7.1in]{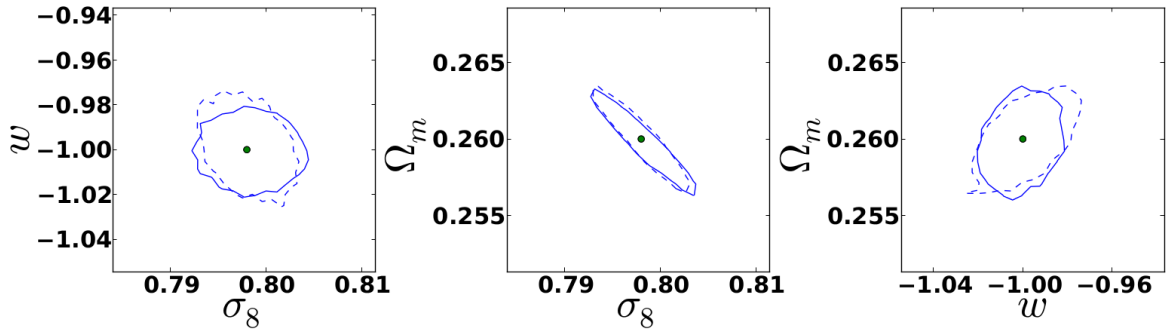}
\caption{68\%\ error contours on the cosmological parameters $\Omega_m$, $\sigma_8$ and $w$ for peak counts, using the forward derivative in the Taylor expansion. The dashed curves show the constraints for measurements including shape noise only, and the solid curves for both shape noise and measurement error. }
\label{fig:contours-fore}
\end{figure*}

Figure~\ref{fig:contours-back} shows the contours for the peak counts above SNR thresholds for the backward derivative. 
The contours in dashed curves show the predicted constraints for measurements made with shape noise only, and solid curves show the constraints for both intrinsic shape noise and the realistic measurement errors described in previous sections. 
The predicted constraints with measurement error  appear at first inspection to be smaller than with shape noise alone. 
In fact this is due to the accuracy with which we can derive the contours from the available set of cosmological simulations. 
A set of simulations covering a larger range of cosmological parameter space would yield smoother contours, but at present this is computationally unfeasible. 
If we look at the contours calculated for the forward derivative shown in Figure~\ref{fig:contours-fore}, we see that in this case the contours with measurement error are roughly the same size as those for shape noise alone. 
The constraints with and without measurement error are  indistinguishable within the statistical accuracy of the contours. 


\begin{figure*}[htb]
\includegraphics[width=7.1in]{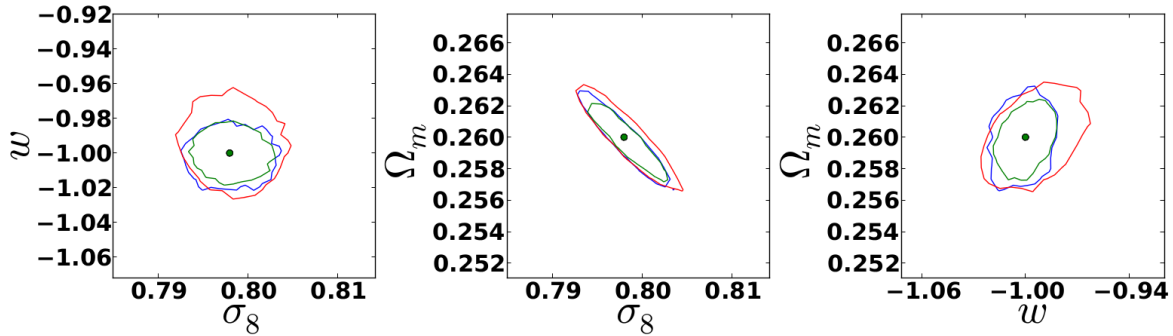}
\caption{68\%\ error contours on the cosmological parameters $\Omega_m$, $\sigma_8$ and $w$ for peak counts (blue) and power spectrum (red) and the combination of the two (green), using the backward derivative in the Taylor expansion  for measurements including shape noise only}. 
\label{fig:contours-combo}
\end{figure*}

This indicates that measurement errors will have a relatively small impact on the accuracy of cosmological constraints with shear peak statistics with LSST. 
We note that we have neglected spatial correlations in measurement errors, which we expect to have a small impact on the number counts of peaks but are known to have a more significant impact on measurements of the power spectrum. 
In comparing constraints obtained with peak counts and our measurement of the power spectrum, we therefore restrict the comparison to the case with shape noise only. 
This is shown in Figure~\ref{fig:contours-combo}. 
As seen in previous work~\citep{Minkowski, Yang:2011zz}, the predicted constraints obtained with peak counts are better than using the power spectrum alone, and a small improvement is found when combining the two measurements.

It should be noted that our finding that measurement errors have a small impact on constraints from peak counts, is only valid under the assumptions that we have made in our analysis framework. 
For example, there may be a larger error associated with the KSB algorithm than we find in this work, because our simulated galaxies are modeled with Sersic profiles rather than real galaxy shapes. 
Since the ImSim input catalogues are anchored to real data, this is a limitation from the current survey data, and will be improved with future observations. 
The accuracy of our constraints is also limited by the accuracy of our N-body simulations, which is discussed in detail in Section V.F of ~\citet{Minkowski}.

The predicted constraints in this work are comparable to those found in other analyses which use the same cosmological simulations, but different measurement techniques~\citep{Minkowski, Yang:2011zz}. 
For example, Figure 12 in~\citet{ Yang:2011zz} shows constraints on cosmological parameters from peak counts from convergence in a single redshift plane, combined with the power spectrum and scaled to an LSST-size survey, that are very similar to our constraints.\footnote{It should be noted that the analysis in ~\citet{ Yang:2011zz} used only 15 galaxies per arcminute$^2$ for one redshift plane, whereas we use 30 galaxies per arcmin$^2$ over the full redshift range.} 
The comparison is not direct, however, due to the differences in analysis methodology. 
In particular, most of the previous work has looked at peaks in maps of convergence, whereas we study reduced shear peak counts. 
The similarity of our constraints to those in \citet{Yang:2011zz}  imply that the difference between peak counts in convergence and reduced shear is small, but we are unaware of any work directly comparing the two methodologies. 

It is even harder to make a comparison with other work that uses an entirely different set of cosmological simulations. 
In that case, the cosmological parameters varied in the simulations, as well as the details of the N-body algorithms, make it almost impossible to make meaningful comparisons between results. 
The best we can say is that the amount of information added by a measurement of peak counts, compared to the power spectrum alone, is consistent with results from other work~\citep{Marian:2010mh, Dietrich:2009jq}.

It should be noted that this work does not use multiple smoothing scales. Using Minkowski functionals as an example, \citet{Minkowski} have shown that combining smoothing scales is important to extract the maximum amount of information from weak lensing maps with a non-Gaussian descriptor, and \citet{Marian:2011rg} have explicitly demonstrated this for peak counts.  
It remains to be studied if the power spectrum contributes any additional information to peak counts when a combination of smoothing scales is used for peaks, or if peaks manage to extract all information when enough smoothing scales are used.  
The work in \citet{Minkowski} noted that combining smoothing scales does not improve the constraints of the power spectrum, because the smallest smoothing scale contains already all of the information contained in the power spectrum. This is true for the Gaussian filter used on convergence in that paper. However, when evaluating aperture mass on reduced shear, one arrives at a compensated filter on convergence. Compensated filters suppress all modes much longer than the size of the filter, in addition to small-scale modes. Thus, for the filter used in this paper---and for using shear in general---combining different smoothing scales could be an asset also for the power spectrum.

%% file: summary.tex
\section{Summary}
\label{sec:summary}

We have produced the first framework for including realistic galaxies and measurement errors in predictions of shear peak counts, using information from the LSST Image Simulator, ImSim. 
Galaxies are drawn from realistic distributions, based on observational data, in redshift, size, magnitude and ellipticity. 
We use information from ImSim to assign uncertainties to the galaxy shape measurements based on these properties, using the KSB shape measurement algorithm. 
We use these realistic galaxies to trace the reduced shear maps produced from ray-traced cosmological N-body simulations, and distort the galaxy shapes appropriately according to the shear parameters interpolated in three dimensions. 
The aperture mass and signal-to-noise ratio is calculated for the resulting simulated catalogues, using an implementation of the aperture mass statistic on the GPU. 

We count peaks above SNR thresholds, and use the resulting peak count distributions to predict constraints on cosmological parameters with LSST. 
We also calculate the aperture mass for an idealized case where we know the reduced shear perfectly, with no uncertainty from galaxy shape noise or measurement error. 
Comparing the two cases, we find that the majority of the discriminating power for the ideal case is in the high SNR peaks, whereas for the realistic measurements the power comes from the full range of peak counts. 
This confirms for the case of peak counts in reduced shear maps what has already been seen in peak counts for convergence maps by ~\citet{Kratochvil:2009wh, Yang:2011zz}---that low and medium significance peaks with SNR$<3.5\sigma$ in reduced shear maps contribute most of the cosmological constraining power.  

We calculate the 68\%\ confidence contours for the  realistic and noiseless peak counts, and find that there is a significant degeneracy in $\Omega_m$ and $\sigma_8$, and smaller degeneracies in the other planes. 
Even in the presence of noise, there is substantial information in peak counts beyond that which can be extracted from the power spectrum alone - the contours from peak counts are approximately 25\% smaller compared to contours from the power spectrum alone. 
Combining the two measurements (from peak counts and the power spectrum), we calculate expected constraints on $\Omega_m$, $\sigma_8$ and $w$ for a ten-year LSST survey of  $0.257 < \Omega_m < 0.263$, $0.792< \sigma_8 <0.804$ and $0.98 < w < 1.02$. 
Note that these constraints are given for illustration, and are not intended as expected final constraints for LSST.  
LSST analyses will use multiple redshifts bins for power spectrum and peaks and include measurements of BAO, supernovae etc. to achieve substantially better constraints.

 We have shown that reduced shear peak counts are a useful probe of cosmology, and that the presence of realistic instrument noise and measurement uncertainties have very little impact on the power of the cosmological constraints. 
This work is the first step in a series of analyses needed to develop the analysis of shear peak counts to the level of sophistication currently enjoyed by the study of shear correlation functions. 
Future work should consider the impact of photometric errors, masked areas, and galaxy clustering around areas of high shear on peak counting, and should start to consider ways to potentially mitigate these effects.